\documentclass[pre,a4paper,oneside,twocolumn]{revtex4}

\usepackage[english]{babel}
\usepackage[latin1]{inputenc}
\usepackage{graphicx}  
\usepackage{latexsym}  
\usepackage{amssymb,amsmath,amsthm,amsfonts} 
\usepackage{color} 
\usepackage{epstopdf}

\begin{document}

\title{First Principles Attempt to Unify some Population Growth Models}


\author{ \firstname{Fabiano}  \surname{L. Ribeiro}}
\email{fribeiro@dfi.ufla.br}

\affiliation{Departamento de F\'isica  (DFI), \\
                Universidade Federal de Lavras (UFLA) \\
	           Caixa Postal 3037 \\
		     37200-000 Lavras, Minas Gerais, Brazil}

\date{\today}

\begin{abstract}
In this work, some phenomenological models, those that are based only on the population information (macroscopic level), are deduced in an intuitive way.  These models, as for instance Verhulst, Gompertz and Bertalanffy models,  are posted in such a manner that all the parameters involved have physical interpretation.
A model based on the interaction (distance dependent) between the individuals (microscopic level) is also presented. This microscopic model reachs the phenomenological models presented as  particular cases. In this approach,  the Verhulst model represents the situation in which all the individuals interact in the same way, regardless of the distance between them. That means Verhulst model is a kind of mean field model. The other phenomenological models are reaching from the microscopic model according to two quantities: i) 
the relation between the way that the interaction decays with the  distance; and ii) the dimension of the spatial structure formed by the population.
This microscopic model allows understanding population growth by first principles, because it predicts that some phenomenological models can be seen as a consequence of the same  individual level kind of interaction.  In this sense, the microscopic model that will be discussed here paves the way to finding universal patterns that are  common to all types of growth, even in systems of very different nature.  
\end{abstract}


\maketitle



\section{Introduction}\label{introduction}

The use of mathematical modeling to understand population growth behaviors has been of great success in the last decades. 
These models find application not only in ecology \cite{Murray2002,Edelstein-Keshet2005,Ribeiro2009}, the immediate aplication, but also in sociology and economy \cite{Ausloos2012,Strzalka2009,Bettencourt2007a}, among other areas of knowledge.
This wide spectrum of applicability of those models has motivated a quest for universal patterns that are present in different types of systems
\cite{Solomon,Cabella2011,Chester2011, Ribeiro2015c, Guiot2003a, West2001, Bettencourt2007a,  Strzalka2008}. 
In the current paper, the inductive process necessary to built population growth models is presented. Next,  an attempt to unify such models by means of a generalized model is also presented. The generalized model is built based on the interaction between the individuals of the population. 
That is, the intention here is to explain  population growth by first principles.

A successful mathematical model must be good enough to predict how the number of individuals of a population behaves, as the time evolves,  according to some ecological properties. 
This should also give some explanation about the properties of the individuals which constitute the population, as the way they interact and how this affects the population as a whole. 

\begin{figure}[t!]
	\begin{center}
		\includegraphics[width=\columnwidth]{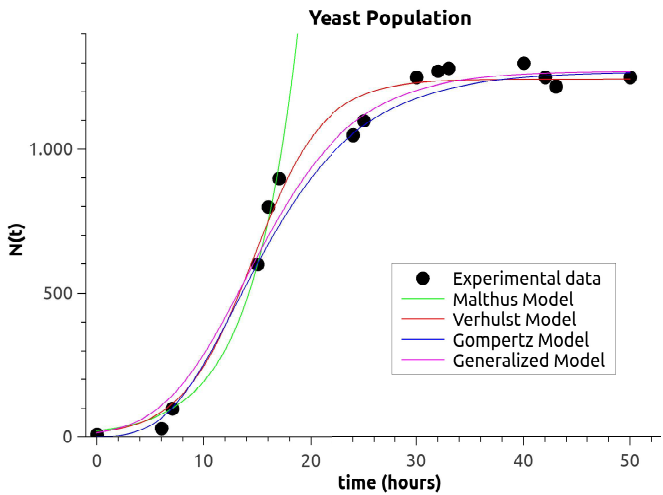}
	\end{center}
	\caption{\label{fig_leveduras} 
	Empirical data (dots) of the growth of a  yeast population. The experimental data were collected from~\cite{Edelstein-Keshet2005}.
	 The solid lines represent the predictions of mathematical models: Malthus, Verhulst, Gompertz, and generalized models.
	 The fit parameter values used are presented in table~(\ref{tabela_valores}). 
	 The Malthus model is good to describe population in the initial stage of growth, but fails for sufficiently large periods.
	 The other models are very similar to describe the empirical data, but the generalized model presents the best fit, with larger  $r^2$. 	 
	}
\end{figure}

A starting point  to build  a population growth model, as all science, is to look at the empirical data. 
The data must give not only the first information about the behavior  of the population, but also the test of the quality of the model.  
If the model is verified as ``good'', then it can be used to formulate hypotheses and some explanation about the system. 
The data of a yeast population growth will be used to illustrate the
creation process of a mathematical growth model and also to test the models.  
Those data, plotted in the figure~(\ref{fig_leveduras}), are the  experimental evidences that allow verifying the validity of the models and inferring some hypotheses.


To organize the ideas from a  mathematical point of view, consider $N(t)$ as the population of yeast (the number or density of individuals) as a function of time $t$. In each time interval $\Delta t$,  the population is updated, which can be written as

\begin{equation}\label{atualiza1}
N(t + \Delta t) = N(t) + \Delta N(t). 
\end{equation}
Here $\Delta N (t)$ is a number representing the balance between the added  (births) and  removed (deaths) individuals in this time interval. In fact, 
$\Delta N(t)$ is the number of births minus the 
number of deaths during the time interval $\Delta t$. 
Of course, if such numbers are known, 
one will know exactly the population size in the next time interval. The problem is that, in general, 
those numbers are not known.
A possible way to solve this problem 
is to extract information from mathematical models.
In the following sections,  two kinds of mathematical models to describe the population dynamics are presented.
The first type, the \textit{ phenomenological models}, or \textit{macroscopic models}, are those that take into account only information of the population as a whole.
Examples of such models are \textit{Malthus, Verhulst, Gompertz} and the \textit{Bertalanffy} models.
The second type, the \textit{microscopic models}, are those that take into account the properties of the system at the individual level. This kind of model 
is built in order to understand the macroscopic phenomena as a consequence, or better,  an  \textit{emergent property},  of the interactions between individuals.
In this way, microscopic models allow to understanding population growth from first principles.

The paper is organized in the following way. 
In section~(\ref{secao_malthus})  the easiest way to represent mathematically the process of growth is presented. The model proposed in this section is the so called Malthus model. 
In section~(\ref{secao_verhulst}), a small corruption in the Malthus model will be done  and then the obtained model, known as Verhulst model,   besides being simple, can  describe the
yeast population growth data very well.  
In section~(\ref{secao_golpertz}), an alternative model, the so-called Gompertz model,
is proposed.  This model also describes the yeast population very well.  
In section~(\ref{secao_generalizado}),  a model that can reach the Malthus, Verhulst and Gompertz models  as a particular case will be proposed, and exactly  because of this,  it is called generalized model. In section~(\ref{secao_micro}), 
the generalized model is deduced by first principle. That is, it is deduced from the interaction between the cells which constitute the population. 
This deduction allows an explanation, at individual (microscopic) level, of the way cells interact to promote the population growth as evidenced by the empirical data (macroscopic level).

\begin{table}
\begin{tabular}{|c|c|c|c|c|c|} 
	\hline Model   &  $q$  &  $N_0$   &  $k_0$  &  $K$ & $r^2$ \\
	\hline Malthus  & 1  & 20  & $0.227$   & $\infty$ & -  \\
	\hline Verhulst & 1 & 15 & 0.302 & 1241 & 0.9865 \\
	\hline Gompertz & $\to 0$ & 1 & 0.1824 & 1268 & 0.9715 \\
	\hline Generalized & 0.394 & 14 & 0.1820  & 1269 & 0.9874 \\ 
	\hline 
\end{tabular}
\caption{\label{tabela_valores}
 Table of the fit parameter values used that best fit the model for the empirical data. 
 The curves generated by these values are presented in  figure~(\ref{fig_leveduras}). The generalized model, as expected, is the best model to describe the empirical data, characterized by the larger $r^2$.} 
\end{table}

\section{Malthus Model}\label{secao_malthus}

In this section,  the simplest way to describe the population growth is considered.  
By sheer simplicity, consider that the number of deaths and the number of births within the population of yeast are proportionate to: i) the number of individuals of the population;  and ii) the time interval. This consideration means that

\begin{equation}
\textrm{Number of Births in $\Delta t$}  =  b  \Delta t N(t),
\end{equation}
and that 

\begin{equation}
\textrm{Number of deaths in$ \Delta t$} =  d \Delta t N(t), 
\end{equation}
where the parameters $b$ and $d$ are constants.
This idea is reasonable because the larger the time interval, the greater the number of births and deaths of individuals is. Likewise, the larger the population,  the greater the number of births and deaths also is. 
This assumption of linearity is quite simplistic,  but will facilitate the analytical development. 

The parameter $b$, the \textit {birth rate},  can be interpreted as the average number of daughter cells generated by each cell of the population per time interval. 
In this simple model which is being proposed, this parameter is considered constant throughout all generations.
The parameter $d$,  the  \textit{mortality rate}, can be interpreted as the proportion of deaths in the time interval. 
Returning to the equation~(\ref{atualiza1}), one has

\begin{equation}\label{atualiza3}
N(t + \Delta t) = N(t) + (b-d) \Delta t  N(t),
\end{equation}
which is a discrete model to describe the dynamics of the population. Given an initial population and the parameters $b$ and $d$,
the size of the population at the time $t$
can be  iteratively calculated. 
When $b>d $, i.e. the birth rate is higher than the mortality rate, then the population grows, otherwise the population decreases.
The values of these parameters can be estimated from the empirical data which one wants to describe. 

The discrete model~(\ref{atualiza3}) can be converted in a continuous equation. It is convenient because it allows analytical analyses,  which  are not possible in the discrete version.
To do it, one takes the limit of infinitesimal time intervals, which means to write

\begin{equation}
\lim_{\Delta t \to 0} \frac{N(t + \Delta t) - N(t)}{\Delta t} = k_0 N(t)
\end{equation}
where $k_0 \equiv b-d$ is  the  \textit{population growth rate}. Identifying the left part of the above equation as a derivative, one has an \textit{ordinary differential equation} (ODE)

\begin{equation}\label{edo_malthus}
\frac{dN}{dt} = k_0 N.
\end{equation}
This ODE can be solved integrating its two sides, which  leads to  the solution

\begin{equation}\label{solucao_malthus}
N(t) = N_0 e^{k_0 t},
\end{equation}
where $N_0$ is the initial population.
Thus, the simple model introduced, that leads to Eq.~(\ref{edo_malthus}),  gives an exponential growth of the population, as suggested by Eq.~(\ref{solucao_malthus}). It is because of this feature that this model is also known as \textit {Malthus model}, referring to the English economist, famous for his studies of populations and the phrase \textit {``The human population grows geometrically while the amount of food increases arithmetically''} \cite{Malthus}.

The problem is that the Malthusian model, precisely because it is very simple, leads to the inconsistency that the population grows indefinitely, blowing up at a sufficiently large time. In fact, this result is inconsistent with what is observed by the data presented in Figure~(\ref{fig_leveduras}), where the yeast population size presents a saturation. The model fails to describe the population as a whole, however,  not everything is lost.
Although the model fails for large times, it is very good to describe the population in the early stages of growth. In Figure~(\ref{fig_leveduras})
one can see that the model fits the empirical data 
very well if one considers the growth in the first 20 hours. This result is quite commendable, given the simplicity of this model.
This fact allows introducing an extremely important concept in modeling, which is the idea of \textit{validity region} of a model. The comparison between the prediction of Malthus model and empirical data presented in Figure~(\ref{fig_leveduras}) clearly shows that the model is good, 
provided it is applied to describe the initial dynamics of the population. Outside this range, the model no longer makes sense.


\section{The Verhulst Model}\label{secao_verhulst}

Essentially, the Malthus model was not good enough to describe the yeast population for a longer time because it does not consider the scarcity of natural resources.
In fact, the malthusian model is too simple, and  therefore one needs to add a little more information to it to get better results.

As the Malthus model behaves well in the early stage of growth, one will still rely on this model, but making a small change in it. Rewriting it  with the introduction of a \textit{correction term} in the ODE~(\ref{edo_malthus}) yields to

\begin{equation}
\frac{dN}{dt} = k_0 N   - \textrm{term}. 
\end{equation}
This ``term'' should be zero (or near zero) when $N$ is sufficiently small (reaching the Malthus model in this regime) and should be maximum when the population reaches  a certain level. Verhulst  \cite{Verhulst1845, Verhulst1847}  considered this corrective term as proportional to $N^2/K $, where $K$ is the \textit {carrying capacity} of the population, which is the maximum size of the population that can be supported by the environment.
Thus the Malthus ODE  with corrective term takes the form

\begin{equation}\label{edo_verhulst}
\frac{dN}{dt} = k_0 N \left[ 1- \left( \frac{N}{K}\right) \right].
\end{equation}
In this model, known as \textit{Verhulst model}, when the population size approaches the carrying capacity ($N \to K$), then $ dN / dt = $ 0, i.e. the population stops growing.
This model predicts that the population saturates when $t \to \infty$,  instead of blowing up, as is the case of the Malthus model.

The solution of this model is obtained by integrating both sides of the differential equation above, resulting in

\begin{equation}\label{solucao_verhulst}
N(t) = \frac{K}{1- \left( 1- \frac{K}{N_0} \right)e^{-k_0 t}}.
\end{equation}

Note from figure~(\ref{fig_leveduras})  that the Verhulst model is very appropriate  to describe the yeast population growth. 
The solution~(\ref{solucao_verhulst})  describes  the population dynamics of yeast very well,  both for early and late time growth. 
One can say that the  Verhulst model, although it is also very simple, with only two parameters (growth rate and carrying capacity), captures \textit{the essence} of the yeast population growth.
This good description happens even disregarding all the details of the interactions between the cells of this population. But then, once you enter with this complexity, it adds more difficulty to solve the problem, and perhaps 
impedes  the analytical treatment, and consequently  disables the understanding of the phenomenon.
With this simple version, considering only phenomenological parameters, it is possible to describe this population  quantitatively, 
conducting to an analytical solution. 
In fact, the model gives a quantitative explanation that the  population of yeast is growing rapidly at the beginning and which has its growth rate decreased as the population is coming to a saturation due to the  scarcity of the environmental resources.

\section{The Gompertz Model}\label{secao_golpertz}

In section~(\ref{secao_malthus}),  it was seen that the malthusian model fails because it leads to an unlimited growth of the population. In section~(\ref{secao_verhulst}),  a way (Verhulst model) 
was  proposed to induce  a saturation in the population dynamics in an attempt to correct the problem of the  Malthus model.
Now, an alternative idea to induce the saturation of the population
is introduced, still using the Malthus model as the starting point.

The indeterminate growth of the Malthus model is due to the per capita growth rate $k_0$ remaining  constant for all individuals of the population and for any generation. A little more information on this growth rate will be introduced now, considering  that this growth rate is time dependent. This allows writing the model~(\ref{edo_malthus}) as

\begin{equation}\label{EDO_gompertz1}
\frac{dN}{dt} = k(t) N.
\end{equation}
To have a situation in which the population stops growing along the time, it is interesting to consider that $k (t)$ decreases with $t$. For example, an exponential decay of this growth rate, i.e. $k(t) = c ^ {- k_0 t}$, where $c$ is a constant and $ k_0 $ plays here the role of a ``half-life'' growth rate, justifying the initial choice in~(\ref{edo_malthus}).
Thus one has a new model

\begin{equation}\label{EDO_gompertz2}
\frac{d N}{ dt} = c e^{- k_0 t} N,
\end{equation}
also known as \textit {Gompertz model} in honor of the English economist who,  in the  XIX century,  used this model to describe human mortality \cite{Gompertz1825} \footnote{The life insurance started to become possible through the  Gompertz ideas in 1825. He realized that the probability of an adult die in the next year increases exponentially with age.
This model is still used by insurance agencies. }.
Integrating both sides of the differential equation~(\ref{EDO_gompertz2}), it is possible to find the solution 

\begin{equation}
N(t) = N_0 e^{- c \left( \frac{1}{k_0}  +  \frac{1}{k} e^{-k_0 t} \right) }.
\end{equation}
Note that for $ t \to \infty $ and $ k_0> 0$ the population converges to $N_0 e^{- c / k_0}$, and therefore this quantity is the very carrying capacity, i.e.

\begin{equation}
K= N_0 e^{-\frac{c}{k_0}}.
\end{equation}
This implies that $ c / k_0 = - \ln ( K / N_0)$,  and consequently the above solution
can be written in terms of the parameters $ K $, $ N_0$ and  $ k_0$, taking the constant $c$ out of the solution. That is, the Gompertz model solution may also be written as

\begin{equation}\label{solucao_gompertz}
N(t) = K e^{\ln\left( \frac{N_0}{K} \right) e^{-k_0 t} }.
\end{equation}

A more convenient way to write the Gompertz model~(\ref{EDO_gompertz2})  is by the ODE

\begin{equation}\label{EDO_gompertz4}
\frac{d N}{ dt} = -k_0 N \ln \left( \frac{N}{K} \right), 
\end{equation}
which is, in fact, more usually found in the literature \cite{Sinai1988, Song2001, Haybittle1998, Molski2008}.
Note that if we integrate the Eq.~(\ref{EDO_gompertz4}),  one has the same solution~(\ref{solucao_gompertz}), which shows that the model~(\ref{EDO_gompertz2})  and the model~(\ref{EDO_gompertz4}) are exactly the same.

Applying the Gompertz model to describe the population of yeast population growth, one gets a result as good as the Verhulst model, as shown in Figure~(\ref{fig_leveduras}). Therefore this model is also interesting to modeling the population dynamics of these microorganisms.

\section{The Generalized Model}\label{secao_generalizado}

It was presented that both the  Verhulst model  (which considers a saturation term) and the Gompertz model (which considers a growth rate that decreases exponentially along the time) lead to satisfactory results to describe the yeast population  growth. 
But if these two essentially different models lead to quite similar results, which of them would be true? Or would both be false?
Upon attempting to answer these questions, it is interesting to work out a more complete model, which has at least the  Verhulst and Gompertz models as special cases.

To do this,  consider that the logarithm function of the Gompertz model~(\ref{EDO_gompertz4}) can be interpreted as a special case of a more general function.
One option would be to look at  the \textit{generalized logarithm function}.
This function is defined by (see Appendix~(\ref{apendice_fgene}) for details)

\begin{equation}\label{eq-ln-q1}
\ln_q (x) \equiv  \frac{x^q - 1}{q},
\end{equation}
where $q$ is the generalization parameter. 
This function is called this way because it recovers the usual logarithm function when one takes the limit $q \to 0$ (see Appendix)
\footnote{Note that $\rm{ln}_{q}(x)$, described from the equation~(\ref{eq-ln-q1}),   \textit {does not} means ``\textit{logarithm of $ x $ in the base $ q $ }''; in this case, one must use the notation $\log_ {q}(x) $.}. 
Replacing, by convenience only,
the usual logarithm function in the Gompertz model~(\ref{EDO_gompertz4})  by the generalized logarithm function, one gets

\begin{equation}\label{eq_modelo_generalizado}
\frac{d N}{ dt} = -k_0  N \ln_q \left( \frac{N}{K} \right), 
\end{equation}
where $ q $ plays the role of a generalization parameter.
This model will be called \textit{generalized model}  because it retrieves not only the Gompertz model  in the limit $ q \to 0$, but also the Verhulst model when $ q = 1 $.
Moreover, if $ q = 1 $ (Verhulst) and $ K \to \infty $, the generalized model also recovers the  Malthus model.
For more details,  see \cite{Cabella2012a, Ribeiro2014, Ribeiro2015b}
The solution of the generalized model can be obtained by integrating both sides of~(\ref{eq_modelo_generalizado}), as usual, which leads to the solution

\begin{equation}\label{Eq_Nt}
N(t) = \frac{K}{ \left[  1- \left[ 1- \left( \frac{K}{N_0} \right)^q  \right] e^{-k_0 t}  \right]^{\frac{1}{q}}   }.
\end{equation}
Note that for $ q = 1 $ and $ q \to 0$, this solution leads to the solutions~(\ref{solucao_verhulst}) and~(\ref{solucao_gompertz}), respectively.


The model was applied to the yeast's experimental data, as can be seen in Figure~(\ref{fig_leveduras}). The best fit was done with $ q = 0.394$, which means that an intermediate model,  between the  Gompertz ($ q \to $ 0) and the  Verhulst ($ q = 1 $) models,  is better to adjust these data.

The model~(\ref{eq_modelo_generalizado})  was originally proposed in \cite{richards1959} and represents an important step in building unified population growth models.
However, the parameter $q$ was introduced here as a 
theoretical argument of generalization, without presenting any physical interpretation.
With a microscopic approach that takes in to account the interactions between individuals, which is presented in Section~(\ref{secao_micro}),  it is possible to obtain this interpretation.
In fact, as discussed just ahead, the  generalization parameter $q$ is related to the interaction field between the individuals of the population.

Besides the model~(\ref{eq_modelo_generalizado}) being based on the macroscopic properties of the system, that is, a phenomenological approach, it can give some suggestions about how the individuals behaves (in average). 
For instance, one can compute, from the model, the typical value of the growth rate of a single individual. That is, the so-called \textit{per capita growth rate}, which will be  represented by $G$. 
This quantity can be computed by (from~(\ref{eq_modelo_generalizado})

\begin{equation}
G(N) = \frac{1}{N} \frac{dN}{dt}  = -k_0 \ln_q \left( \frac{N}{K} \right), 
\end{equation}

\begin{figure}[t!]
	\centering
	\includegraphics[angle=-90,width=\columnwidth]{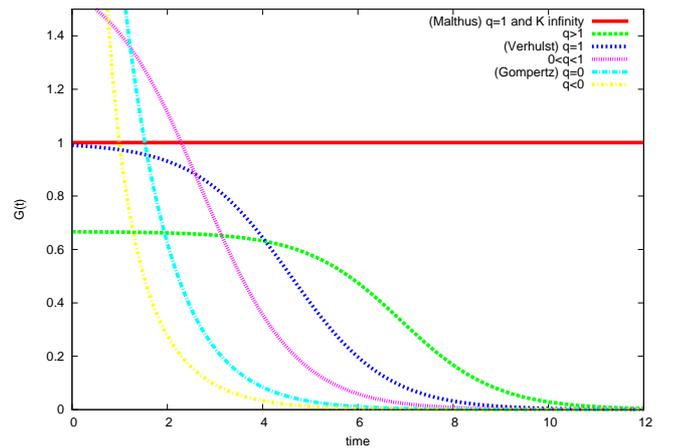} 
	\caption{\label{Fig_Gxt}Graph  of the \textit{per capita} growth rate as a time function.
	Except the Malthus model, in which the \textit{per capita} growth rate is constant throughout the generations, the other particular cases of the generalized models (as Verhulst and Gompertz models) differ only in the way that this average individual growth rate decreases in time.}
\end{figure}

If Eq.~(\ref{Eq_Nt}) is put in the equation above, the time evolution of the \textit{per capita} growth rate,  $G(t)$, is obtained. The figure~(\ref{Fig_Gxt})  presents the time evolution of this quantity. 
With the exception of Malthus model, all the models covered by the generalized model present a \textit{per capita} growth rate which decreases with time. In this way,  Verhulst, Gompertz  and other particular cases differ from one another just in respect to the way that the individuals of the population decrease the capability to reproduce as the time evolves.

It is interesting that $G$ is an individual property that was deduced from the macroscopic information, also known by \textit{top down} approach. This way,  this quantity is not what one can call by ``firt principles'' (\textit{bottom up} approach). 
It will be done in the next section, when one introduces a microscopic model for the interaction between the cells of the population.

\subsection{The Generalized Model and the  von Bertalanffy Model}

The generalized model,  when written in a more explicit form, i.e. using Eq.~(\ref
{eq-ln-q1}), becomes

\begin{equation}
\frac{dN}{dt} = -\left( \frac{k_0}{q K^q}  \right) N^{q+1} + \frac{k_0}{q} N. 
\end{equation}
Or, in a more compact form, one can write it as

\begin{equation}\label{modelo_bertalanfy}
\frac{dN}{dt} = a N^{\beta} - bN,
\end{equation}
where the parameters
 $a \equiv - \frac{k_0}{qK^q}$,  $ b \equiv - \frac{k_0}{q}$, and $\beta \equiv  q + 1$ were introduced. 
The model~(\ref{modelo_bertalanfy}) is known in the literature by \textit {von Bertalanffy model}\cite{Review1949, Kozusko2007}.
This model is often used to describe ontogenetic growth \cite{West2001} and has recently been applied to describe human populations \cite{Bettencourt2007a}.
Thus, the generalized model~(\ref{eq_modelo_generalizado}) is the very von Bertalanffy model. Interestingly, this identification provides   some interpretation for the parameters of the von Bertalanffy model.
That is, by comparison, the model~(\ref{modelo_bertalanfy}) presents:

\begin{itemize}
	\item growth rate:   $k_0 = -b(\beta-1)$;
	
	\item carrying capacity:  
	$K = \left( \frac{b}{a}\right)^{\frac{1}{\beta -1}}$;

	\item generalization constant:  $ q = \beta -1$.

\end{itemize}

In the same way that the generalization model~(\ref{eq_modelo_generalizado}) yields to some phenomenological population growth models as  particular cases, the von Bertalanffy model  also does that. For instance:  
$\beta =2$ ($q=1$) conducts to Verhulst Model if the carrying capacity is finite,  or to the Malthus model otherwise;  and $\beta \to 1$ ($q\to0$) conducts to the Gompertz model. 

However, as there is no physical interpretation of the parameter $q$ yet, there is also no interpretation of $\beta$.
This interpretation will be achieved by a  microscopic model based on the interactions of the individuals, which is presented in the next section.

\section{Microscopic Model}\label{secao_micro}

So far, the presented models have been built  considering only information at the macroscopic level, i.e. populational information.
However, to understand the population dynamics from a deeper point of view,  it is very important to understand how the interactions between individuals perform.
This approach, which can be understood as a microscopic (or \textit{bottom up}) modeling, should lead to understanding the dynamics of the population by first principles.

The first step towards that understanding was obtained from a model presented by Mombach et al in ref. \cite{mombach2002} and reworked in \cite{DOnofrio2009a, Ribeiro2015b, Ribeiro2015c}. This model presents, as result,  the fact that some observed macroscopic properties are a consequence of the inhibitory interactions between the microcomponents of the system. And yet, these researchers succeeded in a physical interpretation of the generalization parameter $ q$.
In this section,  this model is discussed in its details.

To begin with, consider that the replication rate of a cell is regulated by two factors: one is intrinsic to the cell, and another  is related to the interactions of the cell  with the other cells of the population.
This means that the following scheme for a certain cell must be valid:

\begin{center}
	\small{[Replication rate] = [self-replicate stimulus] + 
	
	+ [stimulus from interaction with other cells].}
\end{center}
Following such scheme, and naming $R_i$ the replication rate of the $i$-th cell, the number of daughter  cells generated by $i$ in a time interval $\Delta t $, is

\begin{equation}\label{eq_R2}
\Delta t R_i = G_i  + J I_i.  
\end{equation}
Here, $G_i $ is the intrinsic ability of the cell to self-replicate; $ I_i $ is the  interaction field felt by this cell and caused by the other cells of the population; and $ J $ is the intensity of this interaction field.
If $ J> 0$,  there is cooperation between cells, and if $J<0$, there is competition between them.

The total number of daughter cells added to the population in a certain generation $t$ is $\Delta N (t) = \Delta t \sum_ {i-1}^N R_i$, which leads to a recurrence equation

\begin{equation}\label{eq_nt}
N(t+\Delta t) = N(t) + \Delta t \sum_{i=1}^N R_i.
\end{equation}
Considering infinitesimal time intervals, i.e. $\Delta t \to  0$, the recurrence equation above, together with~(\ref{eq_R2}),  becomes the ODE

\begin{equation}\label{eq_edo_micro}
\frac{dN}{dt} = N \langle G \rangle + J \sum_{i=1}^N I_i.
\end{equation}
where $\langle G \rangle \equiv (1/N)\sum_{i=1}^N G_i$
is the average value of the intrinsic replication capacity of a single cell.

\begin{figure}[t!]
	\centering
	\includegraphics[width=\columnwidth]{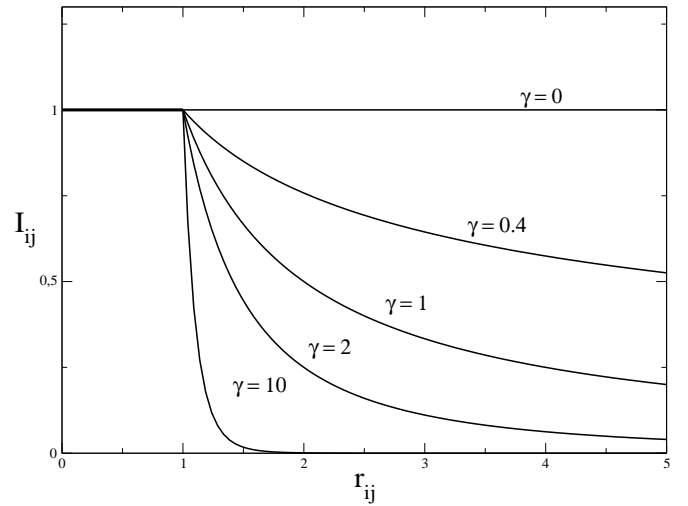} 
	\caption{ \label{figure_decay} 
	Graph representing the interaction field between two individuals ($ i $ and $ j $) depending on the distance $ r_ {ij} $ between them, according to some decay exponent values. It considers $ r_0 = 1/2 $ (fixed).
	The larger $ \gamma $  is, the faster the decay of the field with the distance is.
	When $ \gamma =  0$, the interaction region  is infinite, i.e. the intensity of the interaction does not depend on distance (mean field situation).
	When $\gamma \to \infty $,  the interaction region is null, which means that the cells do not interact.}
\end{figure}

The interaction field  $I_i $ is the result of the interaction of the cell with all the other cells. This allows writing the sum 
$I_i = \sum_{j\ne i} I_{ij} $, where $I_ {ij}$ is the interaction field between the cells $ i $ and $ j $.
Consider that the interaction $I_{ij}$ decays with the distance $ r_ {ij} $ between them according to

\begin{equation}\label{Eq_Iij}
I_{ij} (r_{ij}) = \frac{1}{r_{ij}^{\gamma}}, 
\end{equation}
where $ \gamma $ is the decay exponent of the interaction field.
When $ \gamma =  0$,  the interaction field does not depend on the distance, i.e. the region of interaction between two cells is infinite. This case is called \textit{mean field} situation. 
When $ \gamma \to \infty $, the region of interaction is zero, which means that cells do not interact at all.  To preserve the internal structure of cells, we consider $ r_{ij} \ge 2 r_0 $, where $ 2r_0 $ is the diameter of the cell. Figure~(\ref{figure_decay}) illustrates how the interaction field decays  with the distance for different values of the exponent decay. For a more generic approach about the interaction function~(\ref{Eq_Iij}) see \cite{DosSantos2015a}.

\begin{figure}[t!]
	\centering
	\includegraphics[width=\columnwidth]{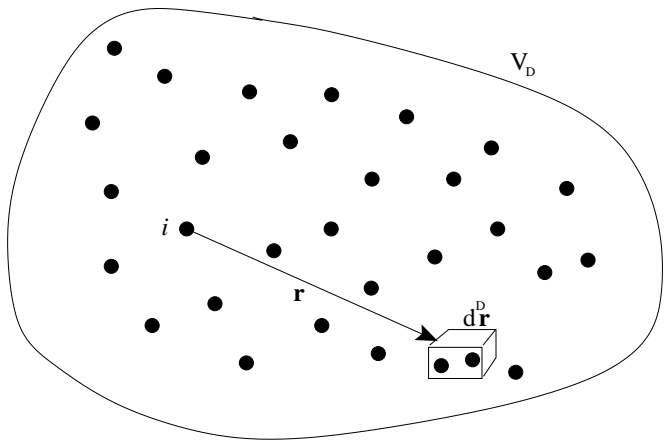} 
	\caption{ \label{fig_estrutura} 
	Hypervolume $ V_D $ in $D$ dimensions, formed by the spatial distribution of the cells (dots). In this figure,  a small group of cells is in a hypervolume element  $d^D\mathbf{r}$,  
	localized by the position vector $\mathbf{r}$,  originated in the $i$-th cell.}
\end{figure}

In relation to the spatial distribution of the cells,  consider that they form  a  structure with hypervolume $ V_D $ in   $ D $ dimensions. For $ D = 3 $ the hypervolume is the usual Euclidean volume; for $ D = 2 $ the population is distributed forming a surface and the hypervolume is the area of this surface; for $D = 1 $ the population is distributed forming a straight segment and, in this case, the hypervolume is the length of this segment.
Consider also $ \rho (\mathbf {r}) $ as the  density of cells in the hypervolume element $ d^D \mathbf {r} $, positioned in $ \mathbf {r}$.
This \textit{position vector}  has its origin in the cell $i$, as described in  figure~(\ref{fig_estrutura}).
The number of cells in this hypervolume element is computed by $dN=\rho (\mathbf {r}) d^D \mathbf {r} $, and if the population is
continuously distributed in the space, the field $ I_i = \sum_ {j \ne i} I_ {ij} $ can be written as the integral $I_i = \int_{V_D} \frac{1}{r^{\gamma}} dN$, which becomes 

\begin{equation}
I_i = \int_{V_D}  \frac{1}{r^{\gamma}} \rho(\mathbf{r}) d^D\mathbf{r}.  
\end{equation}

Assuming, for simplicity only,  that the population is homogeneously distributed in space, i.e.
$\rho(\mathbf{r}) = \rho_0 = \textrm{cte}$, and using hyperspherical coordinates (see Appendix for details), the total field felt by this individual is  

\begin{equation}\label{Eq_Ii}
I_i = I(N) = \frac{\tilde{\rho}_D}{D} \ln_{ 1- \frac{\gamma}{D}  }\left[ \frac{N}{(\tilde{\rho}_D/D)} \right].
\end{equation}
where $\tilde{\rho}_D$ is a parameter related to the density of cells and which depends only on the dimension $D$.
Although it was calculated the interaction field of the specific cell $i$, the result proves identical to all the other cells,  as the right side of the Eq.~(\ref{Eq_Ii}) does not depend on the ``index'' $i$, and consequently it is identical to all the other individuals of the population, regardless of the position of a single cell.  
In fact, the interaction field in a single individual depends only on the size of the population, i.e. $ N$, which justifies to  write $ I_i = I (N) $ in~(\ref{Eq_Ii}).

Back to Eq.~(\ref{eq_edo_micro}), one gets

\begin{equation}\label{resultado-mombach}
\frac{dN}{dt} =  \left[  \frac{J}{(1- \frac{\gamma}{D})} \left( \frac{\tilde{\rho}_D}{D}  \right)^{\frac{\gamma}{D}}  \right] N^{2- \frac{\gamma}{D}}
+  
\left[ \langle G \rangle -  \frac{J \tilde{\rho}_D }{D - \gamma} \right] N
\end{equation}
which is nothing more than the von Bertalanffy growth model given by Eq.~(\ref{modelo_bertalanfy}) (and consequently the generalized model~(\ref{eq_modelo_generalizado}))
with 

\begin{equation}
a=  \frac{J}{(1- \frac{\gamma}{D})} \left( \frac{\tilde{\rho}_D}{D}  \right)^{\frac{\gamma}{D}},
\end{equation}

\begin{equation}
b =   \frac{J \tilde{\rho}_D }{D - \gamma} - \langle G \rangle, 
\end{equation}
and 

\begin{equation}
\beta = 2 - \frac{\gamma}{D}.
\end{equation}
This way, one has here an interpretation for the growth rate $k_0$, the carrying capacity $K$,  and the generalization parameter $ q$ (or $\beta$) from this microscopic model. Actually,  from the above relations, one gets, by a microscopic point of view, the following interpretation for these quantities:

\begin{itemize}

	\item generalization constant:
	
	\begin{equation}\label{q-gamma}
	q = 1- \frac{\gamma}{D};
	\end{equation}

	\item  growth rate:
	
	\begin{equation}\label{Eq_gene_k0}
	k_0 =  \langle G \rangle  q  -  \frac{J \tilde{\rho}_D}{ D};
	\end{equation}

	\item carrying capacity:

	\begin{equation}\label{K-richards}
	K = \left( \frac{ \frac{\tilde{\rho}_D  J}{ D} - (1- \frac{\gamma}{D}) \langle G \rangle }{ \left( \frac{\tilde{\rho}_D}{D} \right)^{ \frac{\gamma}{D} } J }     \right)^{\frac{1}{1-\frac{\gamma}{D}}}.
	\end{equation}

\end{itemize}

The microscopic model allows understanding that the generalization parameter $q$ is given by a relation between the interaction decay among the cells and the dimension of the structure formed by the population.  
This result proves important in the sense that it has only macroscopic quantities, although it has been obtained by microscopic premises.
In other words, the global behavior emerges from local interactions between the cells which compose  the population.
Most importantly, this microscopic model explains, by first principles, all the phenomenological models discussed previously, and gives a physical interpretation of all  quantities involved.

\subsection*{Limit cases: Verhulst and Gompertz Models}

As it was seen in the previous section, the generalized model~(\ref{eq_modelo_generalizado}) has as particular cases the Gompertz and the Verhulst  models.
In the microscopic model, when the dynamic interaction takes place with an  exponent decay $\gamma = 0$, which means that the intensity of interaction between the cells does not depend on the distance, one has  Verhulst dynamics. According to Eq.~(\ref{q-gamma}), one has, in this case,  $q = 1$.
Thus, the Verhulst model can be interpreted as a mean field approach to the population dynamics.
In this limit case, from~(\ref{Eq_gene_k0}) and~(\ref{K-richards}) one has growth rate 
$ k_0  =  \langle G \rangle - J \tilde{\rho}_D/ D$,  and carrying capacity $K = - \tilde{\rho}_D  \langle G \rangle/(JD)$, respectively.

The Gompertz model emerges when  $\gamma \to D$ ($q \to 0$). 
That is, this phenomenological model takes place when the interaction between individuals decays with an exponent equal to the dimension of the system.
Taking the limit  $\gamma \to D$ in~(\ref{Eq_gene_k0}) and~(\ref{K-richards}), 
the population must grow with a growth rate $k_0  =  - J \tilde{\rho}_D/D$, until it reaches the carrying capacity $K = (\tilde{\rho}_D/ D) e^{- \frac{ D \langle G \rangle}   {J \tilde{\rho}_D}}$.

The fact that the microscopic model conducts to the generalized model shows that 
it is possible to find universal patterns in population growth -  
universal here in the sense that properties that are present in all kinds of growth should exist, even in essentially different systems.
In fact, according to the model presented, a subtle difference in the way that the individuals interact leads to a huge difference in the ecological properties. 
This fact can be seen as a motivation to interpret all diversity of growth in nature as a result of a single and universal rule which describes the  way of interaction of the components of the system.

\section{Conclusion}\label{section_conclusion}

In the present work, some important phenomenological population growth models were developed,  with physical interpretation of all the parameters involved.
The models were also compared to empirical data (yeast population growth) to test their validity. 
It was also  shown  that those models can be reached as a particular case of a microscopic model, which takes into account the interaction (depending on the distance) between the individuals of the population.  
From this approach,  it was verified that the Verhulst model represents a situation in which  each individual of the population interacts with all the others in the same way, regardless of the distance that separates them. That means the Verhulst model is a kind of mean field interaction model. Other models, as Gompertz and the generalized one,  are reached according to the relation between the interaction decay exponent and the dimension of the structure formed by the population. 

The phenomenological models presented here can be seen as a consequence of the same individual level kind of interaction.
In this sense, this microscopic model takes an important step towards a more profound understanding of the population growth,  and also  connects many types of growth in a single approach. 
This way, it paves the way to finding universal patterns, common in all types of growth, even in systems of very different nature.

\section*{Acknowledgements}
We would like to acknowledge the useful and stimulating  discussions with Alexandre Souto Martinez.  

\appendix

\section{Generalized Logarithm and Exponential Functions}
\label{apendice_fgene}

The \textit{logarithm function} $\ln(x)$ can be seen, among other interpretations,  as the area below the hyperbole function $f(t)=1/t$, given by the integration

\begin{equation}
\ln(x) = \int_1^x \frac{1}{t} dt.
\end{equation}
We can use this idea to build a \textit{generalized logarithm function}, so that the usual logarithm function is only a particular case.
Defining  such  generalized logarithm function as the area below the unsymmetrical hyperbole $f_{q}(t) = 1/ t^{1- q} $ in the interval $t \in [1,x]$, shows that

\begin{displaymath}
\ln_{q}(x)=  \int_1^x \frac{dt}{t^{1-q}} =
\left\{ \begin{array}{ll}
\frac{x^q -1}{q} & \textrm{for } q \ne 0 \\
 &  \\
\ln(x)  & \textrm{for } q \to 0.
\end{array} \right.
\end{displaymath}

This function is thus a generalization, by introducing the parameter $ q$, of the natural logarithm function.
After all, it retrieves the logarithm function in the limit $q \to 0$, that is

\begin{equation}\label{eq-log-normal}
\lim_{q\to 0} \ln_{q}(x) = \ln(x).
\end{equation}

The inverse function of the generalized logarithm function is the  \textit{generalized exponential function}, defined by

\begin{equation}
e_{q}(x) =   
\left\{ \begin{array}{ll}
\lim_{q^{'} \to q}   (1+ q^{'} x)^{ \frac{1}{ q^{'}} } & \textrm{ if $qx > -1$;} \\
 & \\
0 &   \textrm{otherwise.} 
\end{array} \right. \; 
\label{def-eq}
\end{equation} 
For $q = 0$, one recovers the exponential function:  $e_{0}(x) = e^x$.

These generalized functions have shown to be important because they allow an easy handling of algebraic expressions, besides recovering particular cases \cite{Cabella2011,Ribeiro2015b}.

\section{Calculation of $I_i$}

In this appendix the calculation of the field felt by a single individual is performed. To do this, consider that

\begin{equation}
d^D\mathbf{r} = r^{D-1} dr d\Omega_D, 
\end{equation}
which yields to

\begin{equation}\label{Eq_int_Ii}
I_i = \tilde{\rho}_D \int_{2 r_0}^{R_{max}} r^{-\gamma +D-1} dr 
\end{equation}
where $\tilde{\rho}_D \equiv \rho_0 \int d \Omega_D $
is a parameter related to the cell density which depends on the dinension $D$; and $\int d\Omega_D$ is the solid angle. In fact, for $D=1$, $\int d\Omega_1 = 2 $; for $D=2$,  $\int d\Omega_2 = 2 \pi$; and for $D=3$,  $\int d\Omega_3 = 4\pi$.
The quantity $ R_{max} $ is the maximum distance between two cells.
To solve the integral~(\ref{Eq_int_Ii}), one gets

\begin{equation}
I_i = \frac{\tilde{\rho}_D}{D-\gamma}\left[ R_{max}^{D-\gamma} - (2r_0)^{D-\gamma} \right].
\end{equation}
Considering for sheer convenience, and without loss of generality, that $ r_0 = 1/2 $,
the equation above is simplified to

\begin{equation}\label{eq_resultado_Ii}
I_i = \frac{\tilde{\rho}_D}{D-\gamma}\left[ R_{max}^{D-\gamma} - 1 \right].
\end{equation}
The distance $R_{max}$
can be written in terms of the number of cells  $N$. To do that note that~$N~=~\int_{V_D} \rho (\mathbf{r}) d^{D}\mathbf{r}$, which, given the above considerations, leads to

\begin{equation}
N= \tilde{\rho}_D \int_{0}^{R_{max}} r^{D-1} dr
\end{equation}
and consequently

\begin{equation}
R_{max} = \left( \frac{D N}{\tilde{\rho}_D}  \right)^{\frac{1}{D}}.
\end{equation}
Introducing such results into~(\ref{eq_resultado_Ii}), one gets

\begin{equation}
I_i = I(N) = \frac{\tilde{\rho}_D}{D(1-\frac{\gamma}{D})} \left[ \left( \frac{N}{(\tilde{\rho}_D/D)}  \right)^{1- \frac{\gamma}{D}}  -1\right]
\end{equation}
or, in a more compact form, using the generalized logarithm~(\ref{eq-ln-q1}),

\begin{equation}
I_i = I(N) = \frac{\tilde{\rho}_D}{D} \ln_{ 1- \frac{\gamma}{D}  }\left[ \frac{N}{(\tilde{\rho}_D/D)} \right].
\end{equation}

Although it was calculated the interaction field of the specific cell $i$, the result proves identical to all other cells,  as the right side of the above equation  does not depend on the ``index'' $i$, and consequently it is identical to all the other individuals of the population, regardless of the position of a single cell.  
In fact, the interaction field in a single individual depends only on the size $N$ of the population.

\section*{References}


\end{document}